# Universal scaling behavior of the upper critical field in strained FeSe$_{0.7}$Te$_{0.3}$ thin films

Feifei Yuan*[1,2], Vadim Grinenko*[2,3], Kazumasa Iida[4], Stefan Richter[2,3], Aurimas Pukenas[3], Werner Skrotzki[3], Masahito Sakoda[5], Michio Naito[5], Alberto Sala[6], Marina Putti[6], Aichi Yamashita[7], Yoshihiko Takano[7], Zhixiang Shi*[1], Kornelius Nielsch[2] and Ruben Hühne[2]

[1]School of Physics and Key Laboratory of MEMS of the Ministry of Education, Southeast University, Nanjing 211189, People's Republic of China

[2]Institute for Metallic Materials, IFW Dresden, D-01171 Dresden, Germany

[3]Institute for Solid State and Materials Physics, TU Dresden, 01069 Dresden, Germany

[4]Department of Materials Physics, Graduate School of Engineering, Nagoya University, Nagoya 464-8603, Japan

[5]Department of Applied Physics, Tokyo University of Agriculture and Technology, Koganei, Tokyo 184-8588, Japan

[6]Dipartimento di Fisica, Università di Genova and CNR-SPIN, Via Dodecaneso 33, I-16146 Genova, Italy

[7]National Institute for Materials Science (NIMS), Tsukuba, Ibaraki, 305-0047, Japan

Correspondence should be addressed to F.Y., V.G. and Z.S.

(Email:ytyf0107@163.com, v.grinenko@ifw-dresden.de, zxshi@seu.edu.cn)

**Abstract**

Revealing the universal behaviors of iron-based superconductors (FBS) is important to elucidate the microscopic theory of superconductivity. In this work, we investigate the effect of in-plane strain on the slope of the upper critical field $H_{c2}$ at the superconducting transition temperature $T_c$ (i.e. -d$H_{c2}$/d$T$) for FeSe$_{0.7}$Te$_{0.3}$ thin films. The in-plane strain tunes $T_c$ in a broad range, while the composition and disorder are almost unchanged. We show that -d$H_{c2}$/d$T$ scales linearly with $T_c$, indicating that FeSe$_{0.7}$Te$_{0.3}$ follows the same universal behavior as observed for pnictide FBS. The observed behavior is consistent with a multiband superconductivity paired by interband interaction such as sign change $s_\pm$ superconductivity.

## 1. Introduction

The discovery of superconductivity in the iron oxypnictide LaFeAs(O,F) has triggered a surge of research on Fe-based superconductors (FBS) [1-4]. Systematic studies revealed several universal scaling behaviors for FBS as for example: (1) the Bud'ko-Ni-Canfield (BNC) scaling of the specific heat jump $\triangle C$ at the superconducting transition temperature $T_c$ with $\triangle C \propto T_c^3$ for the majority of FBS [5-11], (2) the linear dependence of the slope of the upper critical field -d$H_{c2}$/d$T$ at $T_c$ versus $T_c$ [12, 13], and (3) the relations between $T_c$ and the structural parameters such as the As–Fe–As bond angle and the anion height [14, 15]. So far, there is no general agreement on the interpretation of these scaling behaviors. Kogan *et al.* proposed that the BNC scaling $\triangle C \propto T_c^3$ and -d$H_{c2}$/d$T \propto T_c$ is related to an intrinsic pair-breaking in superconductors with strongly anisotropic order parameters, such as FBS [12, 16]. Alternatively, Zaanen *et al.* discussed the idea that BNC scaling is expected for a quantum critical metal undergoing a pairing instability [17, 18]. Moreover, Bang *et al.* pointed out that the observed scaling behaviors can be a generic property of the multiband superconducting state paired by a dominant interband interaction [19, 20].

In the case of $FeSe_{1-x}Te_x$, the universal scaling behavior was found for the anion height position [14] and for $\triangle C$ [11]. However, it has not been reported for $H_{c2}$ so far. Recently, we demonstrated that biaxial in-plane strain allows to change $T_c$ of the $FeSe_{1-x}Te_x$ thin films with $x \approx 0.3$ in a broad temperature range avoiding phase separation [21]. This allows to study the behavior of $H_{c2}$ at well-defined conditions. In this work by measuring the electrical resistance in magnetic field we show that also $FeSe_{0.7}Te_{0.3}$ thin films follow the universal scaling behavior of $-dH_{c2}/dT \propto T_c$ in a broad range of $T_c$, as observed in $Ln$OFeAs ($Ln$: lanthanoid elements) and $AEFe_2As_2$ (AE: alkaline earth elements) compounds [12, 13].

## 2. Experiment

The thin films were prepared starting from a stoichiometric $FeSe_{0.5}Te_{0.5}$ target on various substrates, namely $(La_{0.18}Sr_{0.82})(Al_{0.59}Ta_{0.41})O_3$ (LSAT), $CaF_2$-buffered LSAT, and bare $CaF_2$ (001)-oriented single crystalline substrates using pulsed laser deposition (PLD) with a KrF excimer laser (wavelength: 248 nm, repetition rate: 7 Hz) under ultrahigh vacuum (UHV) conditions with a background pressure of $10^{-9}$ mbar [21, 22].

The lattice parameter *a* was derived from reciprocal space maps measured in a PANalytical X'pert Pro system. Transmission electron microscopy (TEM) investigations of the films were performed in a FEI Tecnai-T20 TEM operated at 200 kV acceleration voltage. TEM lamellae were prepared by a focused ion beam technique (FIB) in a FEI Helios 600i using an acceleration voltage of 3 kV in the last FIB step. The composition of the samples was determined by energy-dispersive X-ray spectroscopy (EDX) with an Edax EDAMIII spectrometer in TEM. EDX line scans across the cross-section of the films confirmed the stoichiometry to be homogeneous over the film thickness, as shown in figure A1 in the appendix. It was found that the composition of the films is $FeSe_{0.7}Te_{0.3}$ within the error-bars of the analysis (few percent) for all studied substrates due to the preference of Fe to bond with Se because of the low formation energy [23]. Electrical transport properties were measured in a Physical Property Measurement System [(PPMS) Quantum Design] by a standard four-probe method, for which 4 pins are collinearly aligned along the edge of the film. More details on these structural properties are found in a recent publication of our group [21].

## 3. Results and discussion

### 3.1 Effect of strain on $T_c$

The temperature dependence of the resistance is shown in figure 1 for $FeSe_{0.7}Te_{0.3}$ films grown on different substrates (i.e. bare LSAT, $CaF_2$-buffered LSAT and bare $CaF_2$) measured in magnetic fields up to 9 T for $H \parallel c$. The dashed lines indicate the fit of the normal state just above the superconducting transition temperature $T_c$ by $R(T) = R_0+AT$, where $R_0$ is the resistance extrapolated to $T = 0$ K (i.e. residual resistance) and A is a constant. $T_c$ is defined as 90% of the resistance in the normal state. As shown in figure 1 (a), the lowest $T_c$ of 6.2 K without magnetic field is measured for the films on bare LSAT substrate. $T_c$ is nearly double by employing a 25 nm $CaF_2$ buffer layer (figure 1 (b)). Furthermore, the $FeSe_{0.7}Te_{0.3}$ thin films on bare $CaF_2$ substrate have the highest $T_c \sim 18.1$ K, shown in figure 1 (c) (the data for additional films can be found in figure A3 in the appendix). The applied magnetic field suppresses $T_c$ resulting in a monotonous broadening of the transition attributed to different temperature dependencies of $H_{c2}$ and irreversibility fields ($H_{irr}$). The temperature dependence of the normalized resistance for different films is shown in figure A2. The value of the residual resistivity ratio defined by $R$(300 K)/$R$(20 K) is consistent with the results reported by other groups [24, 25], and is nearly substrate independent.

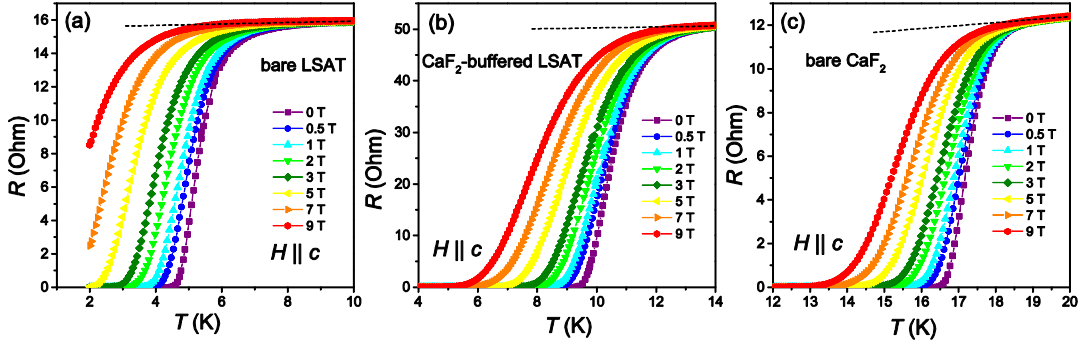

Figure 1. Temperature dependence of the resistance measured in different magnetic fields close to the superconducting transition temperature of the $FeSe_{0.7}Te_{0.3}$ films on (a) bare LSAT, (b) $CaF_2$-buffered LSAT and (c) bare $CaF_2$ substrates in magnetic fields up to 9T for $H \parallel c$. The dashed line indicates the extrapolation of the normal state resistance.

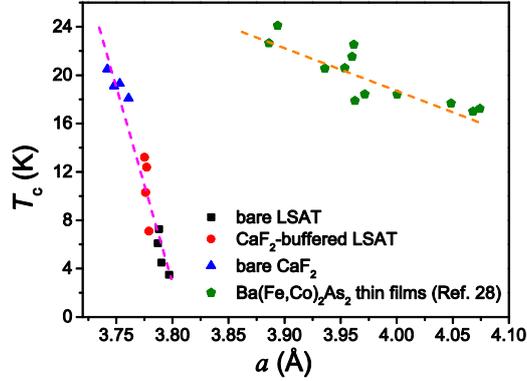

Figure 2. Relation between $T_c$ and the $a$-axis lattice parameter for a series of the $FeSe_{0.7}Te_{0.3}$ films on different substrates. Some of the data are taken from Ref. [21]. The line is a guide for the eye. The data of $Ba(Fe,Co)_2As_2$ thin films from Ref. [28] are also plotted.

As mentioned above, the particularities of the crystal structure have a strong effect on $T_c$ in FBS [26, 27]. Recently we found that $T_c$ of the $FeSe_{0.7}Te_{0.3}$ films is very sensitive to in-plane lattice parameter $a$ [21]. $T_c$ as a function of the $a$-axis is shown in figure 2 for a number of $FeSe_{0.7}Te_{0.3}$ films grown on the mentioned substrates. For comparison, values of optimally doped $Ba(Fe_{0.92}Co_{0.08})_2As_2$ thin films on different substrates are also plotted [28]. The films have different $a$-axis lengths due to different in-plane compressive strain resulting mainly from the large thermal misfit between the substrates and the $FeSe_{1-x}Te_x$ layer [21, 22, 28-32].

It is apparent that the superconducting transition temperature $T_c$ decreases linearly with increasing *a*-axis lattice parameter for the FeSe$_{0.7}$Te$_{0.3}$ films, which is consistent with reports of other groups [33-35]. A linear dependence of $T_c$ on the crystallographic *a*-axis was also found for Ba(Fe,Co)$_2$As$_2$ thin films [28]. However, FeSe$_{0.7}$Te$_{0.3}$ has a much steeper slope. The high sensitivity to strain in the FeSe system can be attributed to the presence of shallow Fermi pockets (with small Fermi energy $\varepsilon_F$) [34]. The strain shifts slightly the position of the bands with respect to the chemical potential resulting in a considerable change of the small $\varepsilon_F$ value or even the appearance of a Lifshitz transition [34]. These changes of the electronic structure can affect $T_c$ as was demonstrated before for the 122 system [30]. This allows us to vary $T_c$ of the FeSe$_{0.7}$Te$_{0.3}$ films significantly solely by in-plane strain.

### 3.2 Upper critical fields near $T_c$

Figure 3 shows $H_{c2}$ for the films on different substrates as a function of $T_c$ for fields parallel to the *c*-axis (the data for additional films can be found in figure A3(e) in the appendix) with 90% criteria for $T_c$. To describe the $H_{c2}$ curves for the FeSe$_{0.7}$Te$_{0.3}$ compound, the spin paramagnetic and the orbital pair-breaking effects should be taken into account [36-38]. A commonly used approach is the WHH theory generalized for multiband superconductors [39]. However, for the reliable determination of the paramagnetic and multiband effects, this analysis requires $H_{c2}$ values in a broad temperature range. Due to very high $H_{c2}$ values of this compound, only the data near $T_c$ are accessible in our experiments. Therefore, we used alternatively an analysis based on the Ginzburg-Landau theory, which provides a simple analytical expression for the temperature dependence of $H_{c2}(T)$ near $T_c$ [40-42]. $H_{c2}$ and its slope near $T_c$ can be estimated by the following equation:

$$H_{c2} = H_{c2}(0) \left[\frac{1-t^2}{1+t^2}\right] \qquad (1)$$

where $t = T/T_c$ is the reduced temperature and $H_{c2}(0)$ is the upper critical field values extrapolated to $T = 0$. We fitted our $H_{c2}$ data using equation (1), as shown by the dashed line in figure 3. The slope $-dH_{c2}/dT$ at $T_c$ is defined using equation (1) at $t = 1$ (figure 4). The analysis including paramagnetic effects results in quantitative changes of the slope (see figure A4 in the appendix) [43]. However, the functional dependence of the slope on $T_c$ is qualitatively unchanged (figure 4 below and figure A4) in spite of relatively strong paramagnetic effects with the Maki parameter $\alpha_M = \sqrt{2}H^*_{c2}/H_p$ given in inset of figure A4 in the appendix, where $H^*_{c2}$ is the orbital limited upper critical field and $H_p$ is the paramagnetic critical field. We found also that $\alpha_M \propto T_c$ in accord with the scaling behavior of $H_{c2}$ discussed in the next section. The large values of the $\alpha_M$ are consistent with previous studies of the FeSe$_{1-x}$Te$_x$ system [44].

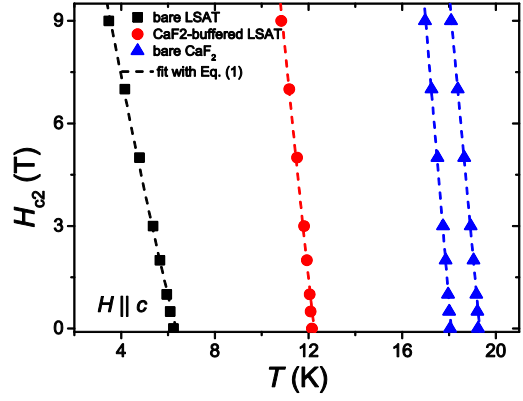

Figure 3. Temperature dependence of the upper critical field $H_{c2}$ with the field applied parallel to the c-axis for FeSe$_{0.7}$Te$_{0.3}$ films with various $T_c$. The dashed lines show the fits based on equation (1).

### 3.3 The slope of $H_{c2}$ at $T_c$ interpreted by a single-band model

The derivative of the upper critical field with temperature -d$H_{c2}$/d$T$ vs. $T_c$ is shown in figure 4. As can be seen, the slope -d$H_{c2}$/d$T$ of FeSe$_{0.7}$Te$_{0.3}$ thin films depends almost linearly on $T_c$ in accordance with the behavior -d$H_{c2}$/d$T \propto T_c$ found in some other pnictides [12, 13]. We note that the choice of criteria does not change the observed linear dependence qualitatively. The -d$H_{c2}$/d$T$ data using 50% of the resistance in the normal state as the criterion of $T_c$ is shown in the inset of figure 4. In contrast to the 90% criterion, the slope does not reach zero by an extrapolation to $T_c = 0$, which indicates that -d$H_{c2}$/d$T$ is affected by additional $T_c$ independent contributions (such as sample inhomogeneity) not related to $H_{c2}$. Therefore, for further analysis we focused on the data obtained with the 90% criterion.

A linear behavior of -d$H_{c2}$/d$T \propto T_c/v_F^2$ is expected for a single-band superconductor in the clean limit, if the Fermi velocity ($v_F$) is the same for different samples [39]. For our samples, the linear relation between $T_c$ and lattice parameters (figure 2) indicates that the mechanism for a $T_c$ suppression is related to the modification of the electronic properties, which should result in a variation of $v_F$ with strain. Therefore, for a clean limit we expect a deviation from the linear behavior -d$H_{c2}$/d$T \propto T_c$. In a dirty limit, the slope -d$H_{c2}$/d$T \propto 1/D$ of an s-wave superconductor is independent on $T_c$, where $D$ is an effective diffusivity constant [45]. It is known that $T_c$ of a single band s-wave superconductor can be suppressed by magnetic impurities. However, smaller $D$ values (stronger impurity scattering) result in a lower $T_c$ and higher -d$H_{c2}$/d$T$ in contrast to the experimental observations. Therefore, for our films effectively a single-band s-wave superconductivity cannot reconcile the whole set of the experimental data.

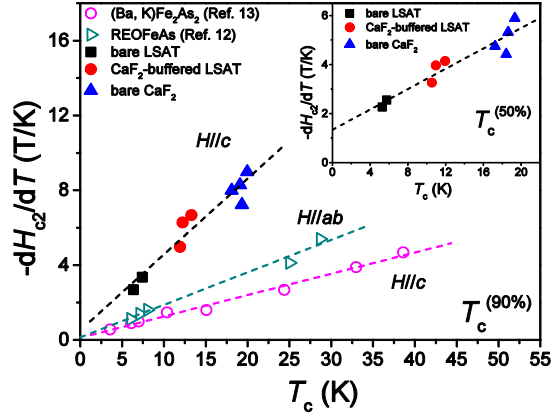

Figure 4. Dependence of the slope -$dH_{c2}/dT$ on $T_c$ for the FeSe$_{0.7}$Te$_{0.3}$ films using 90% criteria for $T_c$. The data obtained using the same criteria for $T_c$ of (Ba,K)Fe$_2$As$_2$ single crystals and REOFeAs are also plotted [12, 13]. The dashed lines are a guide for the eye. Inset: dependence of the slope -$dH_{c2}/dT$ on $T_c$ for the FeSe$_{0.7}$Te$_{0.3}$ films using 50% criteria for $T_c$.

### 3.4 The slope of $H_{c2}$ at $T_c$ interpreted by a two-band model

In the case of superconductivity driven by a single leading interband interaction such as $s_\pm$, the minimal model that describes the system close to $T_c$ is a two-band model [46]. The most of available experimental data are consistent with this picture. So far we have found that only hole overdoped Ba$_{1-x}$K$_x$Fe$_2$As$_2$ system that breaks time reversal symmetry in the superconducting state cannot be described by a two-band model [47]. In this case, a three-band model is needed to describe the superconducting properties [48]. However, general scaling behaviors do not hold in this case (see introduction in Ref. [47]). Therefore, the observed university in scaling behavior of $H_{c2}$ relates our FeSe$_{0.7}$Te$_{0.3}$ films to the majority FBS with superconductivity driven by a single leading interband interaction.

It was shown that the pair-breaking parameters do not alter the slope for dirty superconductors with a sign changed order parameter such as $s_\pm$ superconductors and for two symmetrical bands the slope is -$dH_{c2}/dT \propto T_c/<\Omega^2 v_F^2>$, where $\Omega$ is the variation of the gap along the Fermi surface [12, 16]. The symmetrical case can be excluded due to the expected variation of the Fermi velocity with strain, which results in a deviation from the linear behavior. The symmetrical case is also inconsistent with the available angle-resolved photoemission spectroscopy (ARPES) data [49]. The effect of impurities on a realistic $s_\pm$ superconductor (non-symmetrical bands and non-zero intraband coupling) is rather complex and results for strong enough impurities in a transition to the $s_{++}$ superconducting state [50]. Therefore, an interpretation of the observed linear behavior based on a strong pair breaking effect is also doubtful for the 11 system.

In a clean two band s-wave superconductor, the slope of $H_{c2}$ is defined by a combination of the Fermi velocities and coupling constants for different bands:

$$-\left.\frac{dH_{c2}}{dT}\right|_{T_c} \propto \frac{T_c}{\left(a_1 v_1^2 + a_2 v_2^2\right)} \qquad (2)$$

where $a_1 = \dfrac{\sqrt{(\lambda_{11}-\lambda_{22})^2 + 4\lambda_{12}\lambda_{21}} + \lambda_{11} - \lambda_{22}}{2(\lambda_{11}\lambda_{22} - \lambda_{12}\lambda_{21})}$ and $a_2 = \dfrac{\sqrt{(\lambda_{11}-\lambda_{22})^2 + 4\lambda_{12}\lambda_{21}} - \lambda_{11} + \lambda_{22}}{2(\lambda_{11}\lambda_{22} - \lambda_{12}\lambda_{21})}$ are

constants, which depend on the intraband $\lambda_{11}$, $\lambda_{22}$ and interband $\lambda_{12}$, $\lambda_{21}$ coupling constants [39]. $a_1 \sim a_2$ holds in the case of an extreme $s_\pm$ superconductivity with the dominant interband coupling, which presumably is the case for FeSe$_{1-x}$Te$_x$ [51, 52], and $a_1 \gg a_2$ for an extreme $s_{++}$ case with the dominant intraband coupling. The latter can be excluded based on the arguments for a single-band case since the leading band dominates the superconducting properties close to $T_c$. In the case of strong interband coupling, the universal behavior $-dH_{c2}/dT \propto T_c$ would indicate that the combination $a_1 v_1^2 + a_2 v_2^2$ is nearly strain independent. It is known that the value of the Fermi velocities considerably varies between different bands in the 11 system [49]. Therefore, according to equation (2), $-dH_{c2}/dT$ is dominated by the fastest Fermi velocity assuming sizeable interband coupling, which is expected in the case of the strongly anisotropic sign change superconducting gap [50, 51] or special $s_{++}$ case [53]. The linear scaling indicates that the fastest Fermi velocity is weakly sensitive to strain assuming a weak variation of the coupling constants with strain. In this case, $T_c$ is mainly defined by the band/bands with low Fermi velocities forming small Fermi surface pockets. This is consistent with empirical conclusions based on the ARPES measurements of various pnictides [54]. The universality of the observed scaling $-dH_{c2}/dT \propto T_c$ for different Fe based superconductors imposes constrain on the possible pairing mechanism and indicates a key role of the interband interactions.

## 4. Summary

The superconducting transition temperature of FeSe$_{0.7}$Te$_{0.3}$ films can be significantly modified solely by in-plane biaxial strain. We observed that the slope of the upper critical field of the strained films is proportional to $T_c$ as found for other classes of FBS materials. The behavior observed indicates a striking similarity in the nature of superconducting state between the FeSe$_{1-x}$Te$_x$ system and iron pnictides. This also suggests that the behavior $-dH_{c2}/dT \propto T_c$ may be a generic property of multiband superconductors paired by a dominant interband pairing potential.

**Acknowledgments**

The authors thank S.-L. Drechsler, D. Efremov and A. Maeda for fruitful discussions and M. Kühnel, U. Besold for technical support. The research leading to these results has received funding from the National Science Foundation of China (Grant No. NSFC-U1432135, 11674054 and 11611140101) and Open Partnership Joint Projects of JSPS Bilateral Joint Research Projects (Grant No. 2716G8251b), the JSPS Grant-in-Aid for Scientific Research (B) Grant Number 16H04646 and the DFG funded GRK1621.V. G. is grateful to the DFG (GR 4667) for financial support. The publication of this article was funded by the Open Access Fund of the Leibniz Association.

# Appendix

## A.1. Structural properties

The structural properties and composition of the films were analyzed using transmission electron microscopy (TEM). The data for the two representative films are shown in figure A1 indicating a homogeneous stoichiometry over the film thickness.

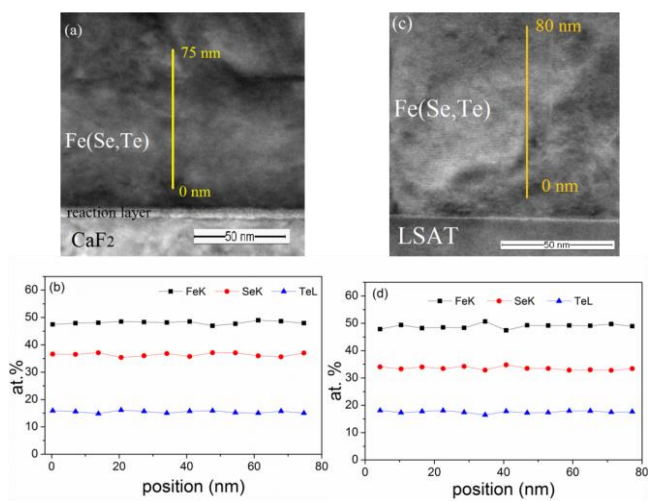

Figure A1. (a) cross-section of the film on bare $CaF_2$. The results for an EDX line scan along the yellow line are shown in (b). The stoichiometry is homogeneous over the film thickness. (c) cross-section of the film on bare LSAT. The results for an EDX line scan along the yellow line are shown in (d). The composition of the films is $FeSe_{0.7}Te_{0.3}$ within the error-bars of the analysis for all studied substrates

## A.2. Electrical resistance

In this section, we provide additional electrical resistivity data (not shown in the main text) measured in zero and applied magnetic field.

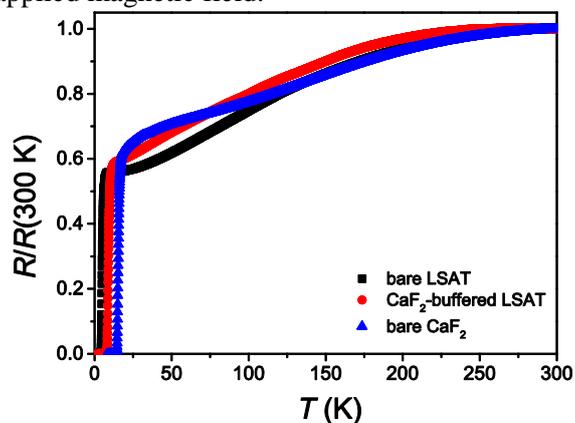

Figure A2. The normalized temperature dependence of the resistance in zero magnetic field for the samples shown in Figure 1 over a large temperature range. The value of the residual resistivity ratio is nearly substrate independent.

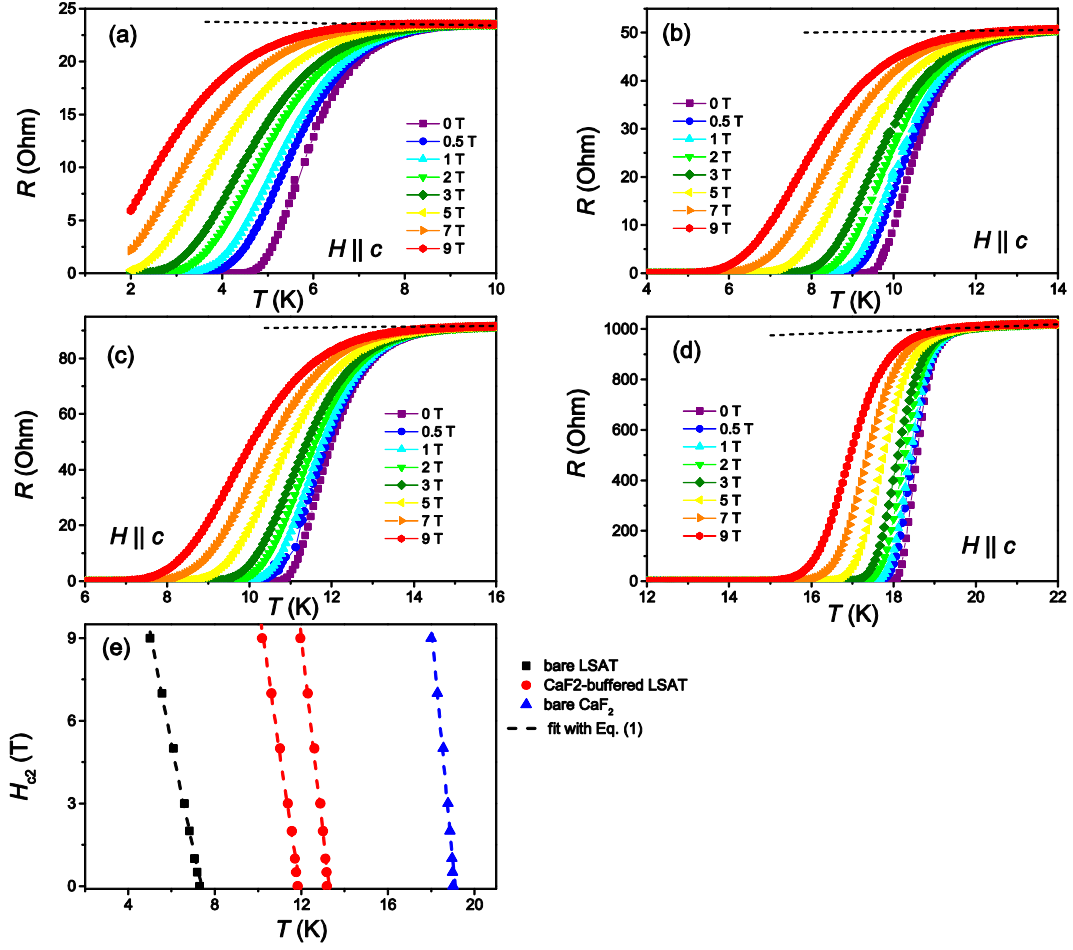

Figure A3. The resistive transition of additional FeSe$_{0.7}$Te$_{0.3}$ films on (a) bare LSAT, (b) and (c) CaF$_2$-buffered LSAT and (d) bare CaF$_2$ in magnetic fields up to 9 T for $H \parallel c$. The dashed line indicates the extrapolation of the normal state resistance. The lowest $T_c$ is measured for the films on bare LSAT substrate. (e) The temperature dependence of the upper critical fields $H_{c2}$ for field parallel to the $c$ axis for FeSe$_{0.7}$Te$_{0.3}$ films with various $T_c$. The dashed line shows the fits based on equation (1) in the main text.

### A.3. $H_{c2}$ analysis

To take into account paramagnetic pair-breaking effects we used an analysis based on the Ginzburg-Landau theory, which provides a simple analytical expression for the temperature dependence of $H_{c2}(T)$ near $T_c$ including paramagnetic effects [43]. As shown by Mineev *et al.*, $H_{c2}$ can be calculated for clean single band superconductors using the following equation:

$$H_{c2} = \frac{e\gamma T_c^2}{a\mu^2}\left[-1 + \sqrt{1 + \frac{\alpha_0 a \mu^2}{(e\gamma T_c)^2}\frac{T_c - T}{T_c}}\right] \quad \text{(A-1)}$$

where $T_c$ is the critical temperature at zero field and $\alpha_0 = N_0$, $a = 7\zeta(3)N_0/4\pi^2$, $\gamma = 7\zeta(3)N_0 v_F^2/32\pi^2 T_c^2$. Here, $N_0$ is the density of states at the Fermi level, $v_F$ is the Fermi velocity and $\mu$ is the magnetic moment.

The slope $-dH_{c2}/dT$ at $T_c$ is given by:

$$-\frac{dH_{c2}}{dT}\Big|_{T_c} = \frac{\alpha_0}{2e\gamma T_c} \quad (A-2)$$

with the Maki parameter defined as:

$$\alpha_M^2 = \frac{\alpha_0 a\mu^2}{(e\gamma T_c)^2} \quad (A-3)$$

Substituting equation (A-2) and (A-3) to equation (A-1) we obtain

$$H_{c2} = \frac{2}{\alpha_M^2}\left(-\frac{dH_{c2}}{dT}\Big|_{T_c}\right)T_c\left[-1 + \sqrt{1 + \alpha_M^2 \frac{T_c-T}{T_c}}\right] \quad (A-4)$$

We fitted our upper critical field data using equation (A-4), as shown by the dash line in figure A4. The obtained slope of the upper critical field -$dH_{c2}/dT$ is shown in figure A4 and the Maki parameter in the inset. Both quantities are proportional to $T_c$ as expected (see also main text) [39].

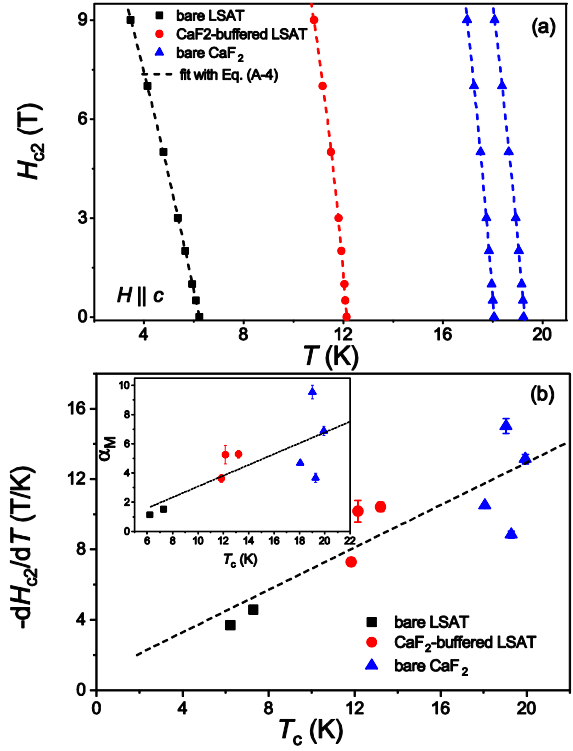

Figure. A4. (a) The temperature dependence of the upper critical fields $H_{c2}$ for fields parallel to the $c$ axis for FeSe$_{0.7}$Te$_{0.3}$ films with various $T_c$. The dashed line shows the fits based on equation (A-4). (b) Dependence of the slope -$dH_{c2}/dT$ on $T_c$ for the FeSe$_{0.7}$Te$_{0.3}$ films obtained by equation (A-4). The dashed lines are a guide for the eyes. Inset: dependence of the Maki parameter $\alpha_M$ on $T_c$. The analysis including paramagnetic effects results in quantitative changes of the slope. However, the functional dependence of the slope on $T_c$ is qualitatively unchanged.

Table 1. Structural and superconducting properties of the films on different substrates presented in this paper.

| Substrate | $a$ (Å) | 90% | | | 50% | | |
|---|---|---|---|---|---|---|---|
| | | $T_c$ (K) | $H_{c2}(0)$ (T) | $-dH_{c2}/dT$ (T/K) | $T_c$ (K) | $H_{c2}(0)$ (T) | $-dH_{c2}/dT$ (T/K) |
| LSAT | 3.787 | 6.35 | 17.11 | 2.69 | 5.27 | 11.99 | 2.27 |
| | 3.788 | 7.39 | 24.83 | 3.36 | 5.74 | 14.66 | 2.55 |
| | 3.777 | 12.23 | 76.79 | 6.28 | 10.96 | 43.47 | 3.96 |
| $CaF_2$-buffer | 3.776 | 11.93 | 59.37 | 4.98 | 10.53 | 34.38 | 3.27 |
| | 3.775 | 13.28 | 88.74 | 6.68 | 11.94 | 49.52 | 4.15 |
| | 3.748 | 19.11 | 158.29 | 8.29 | 18.62 | 98.99 | 5.32 |
| $CaF_2$ | 3.753 | 19.31 | 139.54 | 7.23 | 18.42 | 81.47 | 4.42 |
| | 3.761 | 18.09 | 144.58 | 7.99 | 17.22 | 81.71 | 4.74 |